\documentclass[times, twoside, watermark]{zHenriquesLab-StyleBioRxiv}
\usepackage{blindtext}
\usepackage{lipsum}
\usepackage{booktabs}

\leadauthor{Shuting Xu}

\begin{document}

\title{Solutions for Mitotic Figure Detection and Atypical Classification in MIDOG 2025}
\shorttitle{Solutions for MIDOG 2025}


\author[1,2]{Shuting Xu $\dag$}
\author[1]{Runtong Liu $\dag$} 
\author[1]{Zhixuan Chen}
\author[1]{Junlin Hou}
\author[1]{Hao Chen}

\affil[1]{The Hong Kong University of Science and Technology, Hong Kong, China}
\affil[2]{Nankai University, Tianjin, China}

\maketitle

\begin{abstract}

Deep learning has driven significant advances in mitotic figure analysis within computational pathology. In this paper, we present our approach to the Mitosis Domain Generalization (MIDOG) 2025 Challenge, which consists of two distinct tasks, i.e., mitotic figure detection and atypical mitosis classification.
For the mitotic figure detection task, we propose a two-stage detection–classification framework that first localizes candidate mitotic figures and subsequently refines the predictions using a dedicated classification module.
For the atypical mitosis classification task, we employ an ensemble strategy that integrates predictions from multiple state-of-the-art deep learning architectures to improve robustness and accuracy.
Extensive experiments demonstrate the effectiveness of our proposed methods across both tasks.

\end {abstract}

\begin{keywords}
Mitotic figure detection | Atypical mitosis classification | Model ensemble
\end{keywords}


\footnotetext{$\dag$ These authors contributed equally to this work.}

\section*{Introduction}

Mitotic figures are critical histological markers for assessing tumor proliferation and grading in cancer diagnosis. Their accurate detection is essential for determining tumor aggressiveness and guiding clinical treatment decisions. 
In recent years, deep learning-based methods have significantly advanced the field of medical image analysis \cite{hou2021periphery,hou2022cross,hou2023diabetic,hou2024self,hou2024concept,hou2025qmix}, including pathology image analysis \cite{xu2020data,chen2025segment} and specifically mitotic figure detection \cite{cirecsan2013mitosis,chen2016mitosis,pan2021mitosis}. These advances demonstrate remarkable potential in supporting pathologists through automated and reproducible analysis.

However, several challenges persist in this field. The high morphological variability of mitotic figures across different tissue types, their similarity to impostor cells (e.g., apoptotic cells or lymphocytes). Furthermore, the introduction of atypical mitotic figures, aberrant cell divisions indicative of genomic instability and higher tumor malignancy, adds another layer of complexity. These atypical forms exhibit subtle and irregular morphological features that are difficult to distinguish even for experts.

The MIDOG 2025 challenge provides two dedicated tracks, including mitotic figure detection and atypical mitotic figure classification. In this paper, we propose two distinct deep learning-based approaches tailored to each track. 
For mitotic figure detection, we introduce a two-stage detection–classification framework that first localizes candidate mitotic figures and subsequently applies a dedicated classification module to refine predictions. Our method effectively enhances the robustness against visually similar impostor cells and the overall detection reliability. For atypical mitosis classification, we employ a model ensemble strategy that integrates predictions from multiple state-of-the-art deep learning architectures. Our methods aim to improve generalization across diverse histological appearances and enhance the discriminability between normal and atypical mitotic figures.


\section*{Material and Methods}

\subsection*{Mitotic Figure Detection}
This section details our methodology for the detection of mitotic figures. We first adopt an anchor-free detection network to localize candidate mitotic regions. To further improve discriminability and reduce false positives caused by visually similar impostor cells, we use a secondary classification module that re-evaluates the detected candidates.

\subsubsection*{Dataset}
The MIDOG++ dataset \cite{aubreville2023comprehensive} contains mitosis annotations for 11,937 mitotic figures across 503 individual tumor cases, spanning 7 different tumor types. The MITOS\_WSI\_CCMCT \cite{bertram2019large} dataset includes more than 40,000 annotated mitotic figures from 32 fully annotated whole-slide images (WSIs) of canine cutaneous mast cell tumors. The MITOS\_WSI\_CMC \cite{aubreville2020completely} dataset focuses on canine breast cancer, containing 13,907 mitotic figure annotations across 21 WSIs.

\subsubsection*{Method}
Our approach for mitotic figure detection follows a two-stage pipeline:

\textbf{Stage 1: Candidate Detection.}
We adopt FCOS \cite{tian2019fcos} as the detection model to generate candidate regions. The detector directly predicts object centers, bounding box offsets, and classification scores in an anchor-free manner, which is well-suited for handling the large scale and morphological variability of mitotic figures in histopathology images. This stage is optimized for high sensitivity to ensure most true mitotic figures are captured.

\textbf{Stage 2: Candidate Classification.}
To refine the detector outputs, we introduce a secondary classification module based on ResNet-50 \cite{he2016deep}. Specifically, for each positive candidate, the corresponding image patch is cropped and passed through the classifier. This step is designed to distinguish true mitotic figures from common impostor cells (e.g., apoptotic cells, lymphocytes), thereby reducing false positives.

\textbf{Integration Strategy.}
During inference, the two stages are applied sequentially: the detector proposes candidate bounding boxes, and the classifier re-evaluates each candidate. Only those candidates confirmed by the classifier are retained as final mitotic figure predictions. 

\subsubsection*{Implementation details}
All images were resized to 224$\times$224. For detection, we applied diverse augmentations such as perspective distortion, color jittering, and defocus blur. The models were trained for 150 epochs with a batch
size of 8 using the Adam \cite{kinga2015method} optimizer.
For classification, standard augmentations including random resized cropping, horizontal flipping, and normalization were used, with training for 100 epochs, batch size of 16, using the SGD \cite{robbins1951stochastic} optimizer. F1 score and average precision (AP) were adopted as evaluation metrics for detection, while accuracy, recall, and area under the ROC curve (AUC) were used for classification. All experiments were conducted in PyTorch on a single NVIDIA GeForce RTX 4090 GPU.

\subsection*{Atypical Mitotic Figure Classification}
This section details our methodology for the classification of atypical mitotic figures. We first investigate the performance of several state-of-the-art deep learning architectures as backbone networks. To further enhance predictive robustness and generalization, we employ a model ensemble strategy that aggregates predictions from multiple high-performing models. 

\subsubsection*{Dataset}
The model development and training were conducted primarily on the official MIDOG 2025 Atypical Training Set \cite{weiss_2025_15188326}, supplemented by the AMi-Br dataset \cite{bertram2025histologic}. The primary dataset, MIDOG 2025, provides detailed annotations for atypical mitotic figures across the entire MIDOG++ dataset \cite{aubreville2023comprehensive}, comprising 11,939 mitotic figures extracted from all 7 histological domains. To augment our training data, we incorporated the AMi-Br dataset, which contains subclassification labels (atypical vs. normal) for mitotic figures from the MIDOG 2021 \cite{midog2021} and TUPAC16 \cite{veta2019predicting} challenges. This supplementary dataset provides 3,720 annotations (832 atypical and 2,888 normal), including both original coordinates and extracted patches. The ground truth labels for both datasets are derived from a blinded multi-expert consensus process. No other external datasets were used, and access to the preliminary evaluation data was restricted during the entire development phase.

\subsubsection*{Network architecture}
We evaluate and compare multiple advanced deep learning architectures to identify high-performance feature extractors for discriminating atypical mitotic figures:
\begin{itemize}
    \item \textbf{Vision Transformer (ViT)} \cite{dosovitskiy2020image} processes an image as a sequence of patch tokens via a Transformer encoder. Its self-attention mechanism enables modeling of long-range dependencies and rich global contextual information, which is critical for identifying subtle morphological features.
    \item \textbf{ResNet} \cite{he2016deep} is a pioneering architecture that utilizes residual connections to alleviate the degradation problem in very deep networks. Its widespread adoption and strong performance across diverse vision tasks make it a reliable baseline.
    \item \textbf{DenseNet} \cite{huang2017densely} employs dense connectivity patterns, where each layer receives feature maps from all preceding layers. This design promotes feature reuse, strengthens gradient flow, and yields compact yet highly expressive models.
    \item \textbf{ConvNeXt} \cite{liu2022convnet} is a modernized convolutional network that incorporates design principles from Vision Transformers. It utilizes depthwise convolutions, enlarged kernel sizes, and an inverted bottleneck structure to achieve high accuracy and efficiency. To further enhance its representational capacity, we incorporate a Convolutional Block Attention Module (CBAM) \cite{woo2018cbam} into the ConvNeXt architecture, enabling more refined and attentive feature extraction.
    \item \textbf{EfficientNet} \cite{tan2019efficientnet} uses a compound scaling method to systematically balance network depth, width, and input resolution. It provides a family of models that offer a superior trade-off between model size and accuracy.
\end{itemize}

\subsubsection*{Loss function}
Let $D=\{(x_i,y_i)\}_{i=1}^N$ denote the training dataset of $N$ samples, where $x_i$ is an input image and $y_i$ is the corresponding ground truth label. 
We optimize our models by minimizing the binary cross-entropy loss with logits (BCEWithLogitsLoss). This loss function measures the discrepancy between the ground truth label $y_i$ and the predicted probability $\sigma(z_i)$, where $z_i$ is the model's raw output (logit) for the $i$-th sample. The loss is computed as:
\begin{equation}
    \mathcal{L}_{BCE} = -\frac{1}{N} \sum_{i=1}^{N} [y_i \cdot \log(\sigma(z_i)) + (1 - y_i) \cdot \log(1 - \sigma(z_i))].
\end{equation}
Combining the sigmoid activation $\sigma$ and the cross-entropy loss in a single function provides superior numerical stability during training.

\subsubsection*{Model ensemble}
To improve classification stability and generalization, we employ a model ensemble strategy \cite{hou2021cmc}. Predictions from multiple individually trained models are aggregated to produce the final output. Specifically, we compute the final predictive probability for a test sample by averaging the predicted probabilities from all base models in the ensemble.

\subsubsection*{Implementation details}
All images were resized to 224$\times$224. During training, we applied strong data augmentation techniques, including random cropping, flipping, rotation, color jittering, defocus blur, elastic transformation, and grid distortion. The models were trained for 30 epochs with a batch size of 32 using the Adam \cite{kinga2015method} optimizer. The initial learning rate was set to 3e-5 and decayed following a cosine annealing schedule. Balanced accuracy was adopted as the evaluation metric. All experiments were implemented on the PyTorch platform and executed on one NVIDIA GeForce RTX 4090 GPUs.

\section*{Results}

\subsection*{Mitotic Figure Detection}
In this section, we present the results of our detection-classification framework on mitotic figure detection.

\subsubsection*{Validation of different detection models}
We first compared several detection architectures on the MIDOG++ dataset with a training-validation split of $8:2$. As shown in Table \ref{table:detection_architectures}, FCOS \cite{tian2019fcos} consistently outperformed Faster R-CNN \cite{ren2015faster} and RetinaNet \cite{lin2017focal}, with the best performance obtained when using the ResNeXt101-32x8d as backbone ($F1 = 0.794$, $AP = 0.782$). RetinaNet yielded the weakest results, suggesting that anchor-based methods are less effective for detecting small and morphologically diverse mitotic figures. 

\begin{table}[t]
\centering
\caption{Validation of different detection models in mitotic figure detection dataset.}  
\label{table:detection_architectures}
\begin{tabular}{lcc}
\hline
\textbf{Model} & \textbf{F1} & \textbf{AP} \\  
\hline
FCOS \cite{tian2019fcos} (ResNeXt50\_32x4d) & 0.773 & 0.702 \\
FCOS \cite{tian2019fcos} (ResNeXt101\_32x8d) & \textbf{0.794} & \textbf{0.712} \\
Faster R-CNN \cite{ren2015faster} & 0.763 & 0.715 \\
RetinaNet \cite{lin2017focal} & 0.749 & 0.698 \\
\hline
\end{tabular}
\end{table}

\subsubsection*{Validation with Different Training Datasets}
We evaluated the impact of different training datasets using FCOS  \cite{tian2019fcos} (ResNeXt101-32x8d). CMC \cite{aubreville2020completely} and CMCCT \cite{bertram2019large} were only used for training, while validation was consistently performed on the MIDOG++ \cite{aubreville2023comprehensive} validation split. As shown in Table \ref{table:training_datasets}, combining CMC with MIDOG++ yielded the best performance ($F1 = 0.803$, $AP = 0.842$), slightly better than MIDOG++ alone. This may be due to overlapping categories between CMC and MIDOG++. In contrast, adding CMCCT (canine mast cell tumor) decreased performance, likely because cross-species differences introduced domain shift.

\begin{table}[t]
\centering
\caption{Validation with different training datasets in mitotic figure detection dataset.}
\label{table:training_datasets}
\begin{tabular}{lcc}
\hline
\textbf{Training Dataset} & \textbf{F1} & \textbf{AP} \\
\hline
MIDOG++  & 0.794 & 0.782 \\
CMC + MIDOG++ & \textbf{0.803} & \textbf{0.842} \\
CMCCT  + CMC  + MIDOG++  & 0.783 & 0.796 \\
\hline
\end{tabular}
\end{table}

\subsubsection*{Validation of different classification models}
We extracted $50 \times 50$ patches from MIDOG++ with a 7:1:2 split for training, validation, and testing. Several backbone networks were compared, as shown in Table \ref{table:classification_models}. ResNet50 \cite{he2016deep} and EfficientNetB2 \cite{tan2019efficientnet} achieved strong results of $ACC = 0.8442$, $AUC = 0.9218$ and $ACC = 0.8421$, $AUC = 0.9165$. Vision Transformer (ViT-B) \cite{dosovitskiy2020image} performed less effectively of $ACC = 0.7582$, $AUC = 0.8300$, suggesting convolution-based models are better suited for this task. Finally, an ensemble of ResNet50 and EfficientNetB2 \cite{tan2019efficientnet} with soft voting achieved the best results ($ACC = 0.8602$, $AUC = 0.9307$), demonstrating the advantage of combining complementary classifiers.

\begin{table}[t]
\centering
\caption{Validation of different classification models in mitotic figure detection dataset.}
\label{table:classification_models}
\begin{tabular}{lccc}
\hline
\textbf{Model} & \textbf{ACC} & \textbf{Recall} & \textbf{AUC} \\
\hline
ResNet50 \cite{he2016deep} & 0.8442 & 0.8343 & 0.9218 \\
ResNet101 \cite{he2016deep} & 0.8400 & 0.8308 & 0.9163 \\
EfficientNetB0 \cite{tan2019efficientnet} & 0.8302 & 0.8245 & 0.9093 \\
EfficientNetB1 \cite{tan2019efficientnet} & 0.8242 & 0.8381 & 0.9199 \\
EfficientNetB2 \cite{tan2019efficientnet} & 0.8421 & 0.8378 & 0.9165 \\
ViT-B \cite{dosovitskiy2020image} & 0.7582 & 0.7522 & 0.8300 \\
ResNeXt \cite{xie2017aggregated} & 0.8409 & 0.8313 & 0.9203 \\
ConvNeXt-large \cite{liu2022convnet} & 0.8158 & 0.8214 & 0.9130 \\
Ensemble & \textbf{0.8602} & \textbf{0.8566} & \textbf{0.9307} \\
\hline
\end{tabular}
\end{table}

\subsubsection*{Validation of detection-classification framework}
Using FCOS \cite{tian2019fcos} (ResNeXt101-32x8d) trained on MIDOG++, the baseline detector achieved $F1 = 0.803$ and $AP = 0.842$, as shown in Table \ref{table:det+cls}. With an additional ResNet50 classifier, $F1$ improved to $0.814$, and the ensemble classifier (ResNet50 + EfficientNetB2) further increased performance to $F1 = 0.8432$ and $AP = 0.787$. These results confirm that the cascaded detection–classification framework effectively reduces false positives and enhances overall detection robustness.

\begin{table}[t]
\centering
\caption{Validation of detection- classification framework in mitotic figure detection dataset.}
\label{table:det+cls}
\begin{tabular}{lcc}
\hline
\textbf{Model} & \textbf{F1} & \textbf{AP} \\
\hline
FCOS \cite{tian2019fcos} & 0.803 & 0.842 \\
FCOS \cite{tian2019fcos} + cls (ResNet50 \cite{he2016deep}) & 0.814 & 0.762 \\
FCOS \cite{tian2019fcos} + cls (Ensemble) & \textbf{0.8432} & 0.787 \\
\hline
\end{tabular}
\end{table}


\subsection*{Atypical Mitotic Figure Classification}
This section presents the evaluation of our proposed method for atypical mitotic figure classification. We first report performance comparisons of various deep learning architectures using 4-fold cross-validation on the training set, followed by the results of our ensemble model on the official preliminary evaluation.

\subsubsection*{Cross validation}

\begin{table}[t]
\begin{center}
\caption{Results (mean $\pm$ std) of cross validation in atypical mitotic figure classification dataset.} 
\label{table:classification}
\begin{tabular}{lc}
\toprule
\textbf{Model} & \textbf{Balanced Accuracy}\\
\midrule
Vit\_b\_16 \cite{dosovitskiy2020image} & 0.8129 $\pm$ 0.0055\\
\midrule
ResNet50 \cite{he2016deep} & 0.8469 $\pm$ 0.0144\\
\midrule
DenseNet121 \cite{huang2017densely} & 0.8379 $\pm$ 0.0070\\
DenseNet169 \cite{huang2017densely} & 0.8261 $\pm$ 0.0138\\
\midrule
ConvNeXt\_large \cite{liu2022convnet} & 0.8794 $\pm$ 0.0082\\
ConvNeXt\_base \cite{liu2022convnet} & 0.8707 $\pm$ 0.0059\\
ConvNeXt\_base+CBAM \cite{liu2022convnet,woo2018cbam} & 0.8720 $\pm$ 0.0057\\
\midrule
EfficientNetB0 \cite{tan2019efficientnet} & 0.8462 $\pm$ 0.0040 \\
EfficientNetB1 \cite{tan2019efficientnet} & 0.8325 $\pm$ 0.0000\\
EfficientNetB3 \cite{tan2019efficientnet} & 0.8256 $\pm$ 0.0121\\
EfficientNetB4 \cite{tan2019efficientnet} & 0.8251 $\pm$ 0.0033\\
\bottomrule
\end{tabular}
\end{center}
\end{table}

As illustrated in Table \ref{table:classification}, a comprehensive comparison of various deep learning architectures was conducted for atypical mitotic figure classification using cross-validation. Among all evaluated models, the ConvNeXt family demonstrates superior performance. Specifically, ConvNeXt\_large achieves the highest balanced accuracy of 0.8794, followed closely by ConvNeXt\_base at 0.8707. The incorporation of the CBAM attention module into ConvNeXt\_base yields a further improvement, achieving an accuracy of 0.8720.
ResNet50 and DenseNet121 establish strong baselines with accuracies of 0.8469 and 0.8379, respectively, confirming the effectiveness of residual and dense connections.
Within the EfficientNet family, a clear trend emerges where smaller models (B0 and B1) outperform their larger counterparts (B3 and B4), suggesting that a more compact model capacity is sufficient for effective feature extraction in this task. The Vision Transformer (ViT\_b\_16), while achieving competitive results, is outperformed by other convolutional architectures. This performance gap may be attributed to the limited amount of training data, which might be insufficient for ViT to fully leverage their representational capacity and achieve optimal performance.

\subsubsection*{Preliminary evaluation}
We ensemble the family of ConvNeXt models to construct our final model. On the preliminary evaluation set of the Atypical Mitotic Figure Classification Challenge, our approach achieves a Balanced Accuracy of 0.86. This result demonstrate the model's robust capability in effectively distinguishing between typical and atypical mitotic figures.


\section*{Conclusion}
In this work, we presented deep learning-based approach to address the two tasks of mitotic figure detection and atypical mitosis classification on MIDOG 2025 Challenge. For mitotic figure detection, our two-stage detection-classification framework effectively localized candidate figures, achieving an F1 score of 0.8432 on the validation set. For atypical mitosis classification, our ensemble strategy successfully integrated predictions from multiple architectures, reaching a balanced accuracy of 0.8794 in cross-validation
These results highlight the potential of our approaches to advance computational pathology applications and contribute to more reliable mitosis analysis across diverse clinical settings.


\section*{Bibliography}
\bibliography{literature}

\end{document}